\documentclass{article}
\usepackage{spconf,amsmath,graphicx}
\usepackage{url}
\usepackage{amsmath}
\usepackage{amssymb}
\usepackage{enumitem}
\usepackage{balance}

\title{SG-VAD: Stochastic Gates Based Speech Activity Detection}

\name{Jonathan Svirsky, Ofir Lindenbaum}

\address{Faculty of Engineering, Bar-Ilan University, Ramat-Gan, 5290002, Israel\\
{\{svirskj, ofir.lindenbaum}\}@biu.ac.il}

\begin{document}

\maketitle

\begin{abstract}
We propose a novel voice activity detection (VAD) model in a low-resource environment. Our key idea is to model VAD as a denoising task, and construct a network that is designed to identify nuisance features for a speech classification task. We train the model to simultaneously identify irrelevant features while predicting the type of speech event. Our model contains only 7.8K parameters, outperforms the previously proposed methods on the AVA-Speech evaluation set, and provides comparative results on the HAVIC dataset. 
We present its architecture, experimental results, and ablation study on the model's components. We publish the code and the models here \textit{https://www.github.com/jsvir/vad}. 
\end{abstract}

\begin{keywords}
Voice Activity Detection, Feature Selection, Speech Recognition\end{keywords}
\section{Introduction}
\label{sec:intro}

Speech activity detection (or Voice Activity Detection - VAD) is crucial in the voice processing pipeline. VAD generally serves as a filtering submodule for the downstream tasks, which are typically much more computationally expensive. A good VAD model should have a small number of parameters and lead to high detection accuracy. It is well-known that there is a trade-off between both of the desired qualities when using deep learning models for VAD. Recently proposed architectures for VAD \cite{kopuklu2022resectnet, jia2021marblenet} presented a breakthrough by improving both the accuracy and reducing the model's size. Kopukli et al. \cite{kopuklu2022resectnet} proposed to use sinc function-based convolution and frequency shift modules to reduce the model size while preserving high accuracy in voice detection. Jia et al. \cite{jia2021marblenet} exploit 1D time-channel separable convolutions to treat the size-accuracy trade-off. Most of the recent works on VAD operate on single frame-level chunks and require postprocessing to aggregate several frame-level predictions into a single output \cite{kopuklu2022resectnet, jia2021marblenet, rho2022vad, braun2021training, xu2021lightweight, xiong2021computationally, kim2022ada, xu2020polishing, makishima2021enrollment}. Inspired by \cite{chen2020voice, lee2020dual, li2022voice} where a segment-level model is also suggested, we proposed a VAD model for segment-level voice detection, hence, a broader input context is provided but not necessarily required. Such a setup removes the need for postprocessing which is required in frame-level VAD models.
We propose a novel model and training regime that reduces the number of parameters of the model and improves the model's accuracy compared with leading baselines.

Intuitively, our framework relies on modeling the VAD problem as a denoising task, where the model classifies between speech and noise. The signal could be represented as a union of disjoint speech events $x= \bigcup{s_{\ell}}$ where $s_{\ell} \in \{noise, speech \}$ at time frame $i$. We have no assumptions on the duration of a single frame. Thus, we can use a short or long segment to represent the signal in the Fourier domain. Feature selection aims at attenuating noisy or nuisance features that are useless for the main prediction task. VAD model could
be trained as a feature selection model that disregards all time frames associated with noise or non-speech. We propose a novel method that exploits the feature selection paradigm to train a VAD model. 
Specifically, we apply the Locally Stochastic Gates (LSTG) mechanism, recently proposed for feature selection \cite{lindenbaum2021differentiable, yang2022locally}. 

VAD model is assumed as Neural Network that predicts the \textit{gates} on the input signal to produce a sparse output for the downstream model. This NN-based gating model is trained with Gaussian-based relaxation of Bernoulli variables, termed Stochastic Gates (STG) \cite{yamada2020feature}, which relies
on the reparameterization trick \cite{miller2017reducing, figurnov2018implicit} to reduce the variance of the gradient estimates. Since the gates are learned as a function of input samples, we denote them as \textit{local} \cite{yang2022locally} in contrast to the global setup \cite{lindenbaum2021differentiable}. By applying LSTG method, we select the most informative features in spectrum representation in both time and frequency dimensions; then, we aggregate the selection results to produce voice activity estimation for an input audio segment.

Specifically, this paper makes the following contributions: 

1. We propose a novel SG-VAD model based on 1D time-channel separable convolutions and the Locally Stochastic Gates mechanism that predicts labels for multiple frames in the audio signal.

2.  SG-VAD achieves state-of-the-art performance on the AVA-speech \cite{chaudhuri2018ava} and HAVIC datasets \cite{strassel2012creating} with minimal training setup.

3.  SG-VAD has 11x fewer parameters than MarbleNet model \cite{chaudhuri2018ava}, and its size is very close to ResectNet model \cite{kopuklu2022resectnet}, which makes it applicable to run on edge devices.

\section{Method Description}
In the following subsections, we detail the training and inference steps of our method and the two main components of our neural network.
\subsection{Method overview}

To describe our method, we distinguish between training and inference. 

\begin{figure}
  \centering
    \includegraphics[width=1.\columnwidth]{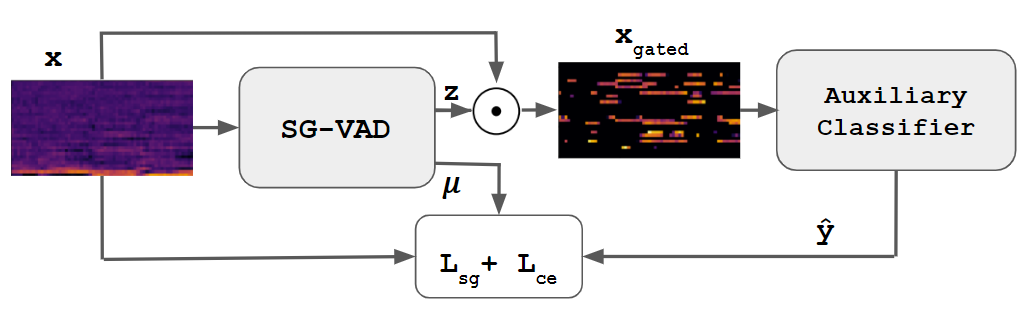}
    \caption{\textit{Training flow of the proposed framework. An auxiliary classifier is used to train the main module SG-VAD. SG-VAD is a dynamic feature selection model that gates uninformative features for the auxiliary classifier. Both models are trained simultaneously by minimizing classification error while selecting a few features. }}
    \label{fig:train}
         \vskip -0.2 in
\end{figure}

\textbf{Training} \hspace{5pt} During the training, which is illustrated in Figure \ref{fig:train}, we train two neural networks in an end-to-end fashion: the first one is the main model SG-VAD which serves as a feature selector model, and the second is an auxiliary module which is trained as a multi-label classifier. To train these modules, we minimize the sum of two loss terms: $L_{sg}$ and $L_{ce}$, which are described in the following sections in more detail. SG-VAD produces the same size output $z$ as its input $x$, a tensor of binary labels. Then the input $x$ is multiplied in an element-wise way by $z$ to produce $x_{gated}$. The classifier, in turn, outputs prediction $\hat{y}$, which is trained to agree with the target class of $x$.

\textbf{Inference} \hspace{5pt} At inference, only the model SG-VAD is used. For the given input tensor, the model outputs the same size tensor with gates on features. The gates are then used to estimate a voice activity label. In the following section, we explain how is it done.

\begin{figure}[t]
  \centering
    \includegraphics[width=.7\columnwidth]{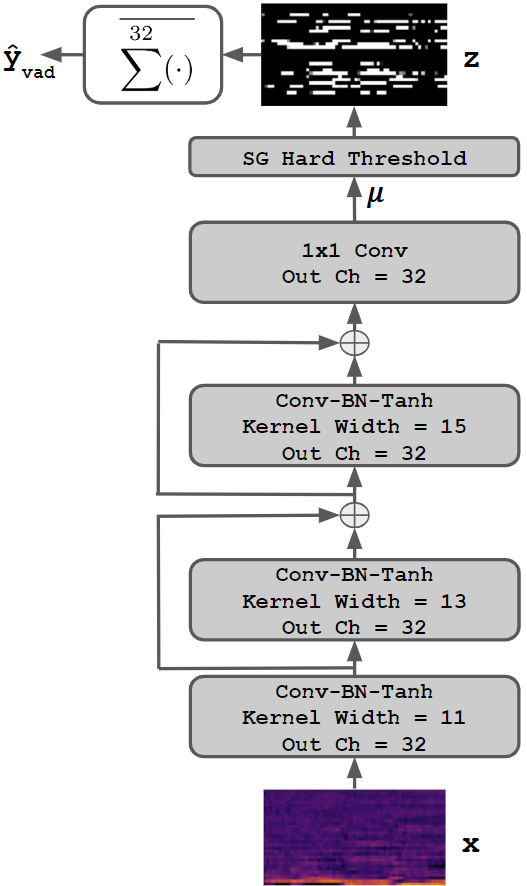}
    \vskip -0.1 in
    \caption{\textit{A diagram of the SG-VAD model. MFCC features are forwarded through residual 1D time-channel separable convolutions with batch normalization and a tanh activation. Hard thresholding is applied to produce binary features $z$, which are aggregated into an output voice activity prediction score $\hat{y}_{vad}$.}}
    \label{fig:proposed}
     \vskip -0.2 in
\end{figure}

\subsection{Stochastic Gates Based VAD (SG-VAD)}
\label{sec:sgvad}

The main module, SG-VAD, is based on 1D time-channel separable convolutions and inspired by the MarbleNet model \cite{jia2021marblenet}. We present its architecture in Figure \ref{fig:proposed}. It includes a 1D time-channel separable convolution layer followed by batch normalization and Tanh activation. Next, two residual layers with the same architecture but different kernel widths are applied. An additional single $1\times1$ convolution layer is applied before the final Stochastic Gates thresholding layer.
During the training, the model accepts extracted features from the audio segment with channel size 32 and variable time dimension and outputs the gated version of the features where nuisance features are replaced with zero values. Assuming that the output of the $1\times1$ convolution layer is $\mu_i$, the model is trained with the loss term $L_{sg}(z_i)=||z_i||_0$ where $z_i$ is denoted as input \textit{gates} and is defined by the hard thresholding function (\ref{eq:gates}):

\begin{equation}\label{eq:gates}
 z_i = max(0, min(1, 0.5 + \mu_i + \epsilon_i)),
\end{equation}
where $\epsilon_i$ is drawn from $N(0, \sigma^2)$ and $\sigma=0.5$ is fixed throughout training. To encourage the model to attenuate the background noises, we calculate and optimize $L_{sg}$ only for samples labeled as \textit{background}. In addition, to provide a gated input to the downstream classifier during the training, $x_i$ is multiplied in an element-wise fashion by $z_i$ to produce the sparse version of $x_i$ which is $x_{gated}=x_i \odot z_i$.

During inference, the VAD label $\hat{y}_{vad}$ for an input $x_i$ with $T$ time frames is estimated by formula (\ref{eq:scores}), where $z_i^j \in \{0,1\}$:

\begin{equation}\label{eq:scores}
 \hat{y}_{vad}(x)= \frac{1}{T} \sum_{i=1}^{T} \sum_{j=1}^{32} z_i^j
\end{equation}

\subsection{Auxiliary Classifier}
\label{sec:aclass}
The second component of our model is a Convnet-based classifier. Specifically, we use MarbleNet3x1x64 \cite{jia2021marblenet} with three blocks without repetitions and a channel size of 64 in each block. The module is trained as a multi-class classifier with a cross-entropy loss $L_{ce}$ function on gated inputs $x_{gated}$ produced by the SG-VAD module. The target labels include different speech event classes and a single class for background noise or non-speech events.
 \begin{equation}\label{eq:loss}
     L(x_i) = \begin{cases}
        L_{sg}(x_i) + L_{ce}(x_i), & \text{if  $y_i=0$}\\
        L_{ce}(x_i), & \text{otherwise}
    \end{cases}
\end{equation}
We train both modules end-to-end to minimize the loss in (\ref{eq:loss}). To discriminate between multiple classes, we propagate the gradients from the cross-entropy loss back to SG-VAD. This way, the SG-VAD model can distinguish between noise and speech more precisely. In addition, the predictions of SG-VAD become more accurate since the STGs attenuate the noise during the training process of the downstream classifier. In section \ref{sec:results}, we show that although the model is trained with minimal training setup compared with previous works, it still outperforms both MarbleNet \cite{jia2021marblenet} and ResectNet \cite{kopuklu2022resectnet} models on AVA-Speech \cite{chaudhuri2018ava} evaluation dataset.

\section{EXPERIMENTS}
\label{sec:exps}

\subsection{Training methodology}
We train our model on a smaller dataset compared to most existing baselines. This can allow us to highlight to algorithmic advantages of our framework. We use Speech Google Commands V2 dataset (GSCV2) \cite{warden2018speech}. The data has 35 classes of speech events and is extended by an additional \textit{background} class. Following \cite{jia2021marblenet}, we include 35 background categories of noises, in total about 2,100 variable-length audio segments, such as "motorcycle", "Bus" and "Static" from \cite{font2013freesound} (FS2K). The samples were obtained using the scripts provided by NeMo \cite{kuchaiev2019nemo}. We split the segments to the maximal length of 0.63 seconds and added them as an additional category to the Speech Google Commands V2 dataset. 
We split the dataset into train/validation parts and chose the best model based on minimal validation accuracy.
Although the dataset includes about ~23 hours of clean speech events which is $\times$20 less than 500 hours in Interspeech 2020 DNSChallenge dataset \cite{reddy2020interspeech} used for ResectNet \cite{kopuklu2022resectnet} training, the model still produces comparable results on the HAVIC evaluation dataset and outperforms all previous methods on AVA-Speech test.

The audio segments are processed by extracting 32 Mel-frequency cepstral coefficients (MFCC) features. The input was augmented with time shift perturbations in the range of T = $[-5; 5]$ ms and white noise of magnitude $[-90;-46]$ dB with a probability of $80\%$. Additionally, SpecAugment \cite{park2019specaugment} was applied with 2 continuous time masks of size $[0; 25]$ time steps, and 2 continuous frequency masks of size $[0; 15]$
frequency bands. SpecCutout \cite{devries2017improved} was also used with five rectangular masks in the time, and frequency dimensions as in \cite{jia2021marblenet}. The model was trained with the SGD optimizer with momentum = $0.9$ and weight decay = $1e-3$. We utilized the Warmup-Hold-Decay learning rate schedule \cite{he2019bag} with a warm-up ratio of 5\%, a hold ratio of 45\%, and a polynomial (2nd order) decay for the remaining 50\% of the schedule. A maximum 
learning rate of $1e-2$ and a minimum learning rate of $1e-4$ were used. We trained all models for 150 epochs on a single NVIDIA GeForce GTX 1080 Ti with a batch size of 128. The model was implemented and trained with NeMo \cite{kuchaiev2019nemo}. 

\subsection{Evaluation Method}
\label{sec:evaluation}

The performance of SG-VAD was evaluated on the AVA speech \cite{chaudhuri2018ava} and the HAVIC \cite{strassel2012creating} datasets. AVA speech
contains manually annotated 15-minute-long clips from 160 YouTube videos. We use the same subset of 122 out of 160 videos as authors in \cite{jia2021marblenet}, in total 30 hours of playback time. The HAVIC dataset contains
72h of audio collected from YouTube videos with manually annotated speech, music, noise, and singing segments. We used all speech-related segments as a target. Since our model is intended to predict speech/non-speech for the full audio segment and not only for a single frame, we are not required to apply any postprocessing procedure and the predictions produced for full audio segments of a clip accordingly to the timing provided in the evaluation set for each segment. An exception is made for too-long chunks that are split into shorter, up to 100 seconds segments.
To obtain results for the MarbleNet model on the HAVIC dataset, we follow the postprocessing and inference description in \cite{jia2021marblenet}. To evaluate performance,
we used the Area Under Curve (AUC) of the Receiver Operating Characteristic (ROC), denoted as AUC-ROC, which is a calibration-free measure of detection performance. 

\subsection{Results}
\label{sec:results}

In table \ref{tab:results}, we provide our evaluation results. 
We distinguish between models by the train datasets used to train them: Kopuklu et al., and Broaun et al. train their models on DNS \cite{reddy2020interspeech} dataset. Rho et al. presented the best results on the Common Voice \cite{ardila2019common} subset with about 200 hours of speech and Audioset subset with noise events \cite{reddy2021interspeech, gemmeke2017audio}. Kim et al. exploits TIMIT dataset \cite{zue1990speech} augmented with noises.

\begin{table}[h!]
  \centering
  \small{
    \begin{tabular}{|c|c|c|c|c|}
    \hline
        Model & Params & Training & \multicolumn{2}{c|}{AUC-ROC}  \\
    
    & & Dataset & AVA & HAVIC\\
        \hline
       \hline
     ResectNet 0.5x \cite{kopuklu2022resectnet}& 4.5k & DNS  & 88.6 &  83.5 \\
    ResectNet 1.0x \cite{kopuklu2022resectnet} & 11.1k &  & 90.0 &  84.9 \\
    Braun et al. \cite{braun2021training} & 1773k &  & 92.4 & 86.8 \\
    \hline
    NAS-VAD \cite{rho2022vad} & 151k & CV, 
    AS & 90.5 & - \\
    \hline
    ADA-VAD \cite{kim2022ada}  & - & TIMIT  & 85.3 & - \\
    \hline
     MarbleNet \cite{jia2021marblenet} & 88k & GSCV2,FS2K & 85.8 & 80.4 \\
    \textbf{SG-VAD} (ours) & \textbf{7.8k} & & \textbf{94.3} & \textbf{83.3} \\
    \hline
    \end{tabular}
    }
  \caption{\textit{Our model significantly outperforms all recently proposed models on AVA corpus. Furthermore, our model trained on a smaller dataset outperforms ResectNet on AVA and produces comparable results on HAVIC dataset}} 
   \label{tab:results}
        \vskip -0.2 in
\end{table}

\subsection{Ablation Study}
\label{sec:ablation}

We provide an ablation study on our model's architectural and loss components. In the first setting (SG-VAD-R), we use only the SG-VAD module and train it as a regression model with mean squared error loss measured between the model output and target label. We intentionally do not add an extra output layer with Softmax normalization to follow the inference setup of SG-VAD. The prediction is based on the summing of learned gates for each time frame.
We add the auxiliary classifier (AC) in the second ablation version, but the training is done without $L_{sg}$ loss term. The third setup includes the $L_{sg}$ term but without supervision on the SG-VAD module: instead of using $L$ in (\ref{eq:loss}) we minimize the sum $L_{sg}(x_i) + L_{ce}(x_i)$ for all $x_i \in X$. Finally, our full model result appears in the last row. From the ablation results in Table \ref{tab:ablation} we can observe the indispensability of all model parts. Once we add the second classifier module without considering all proposed loss terms (SG-VAD + AC), the model performance degrades even lower than a deep regression model (SG-VAD-R). In addition, SG-VAD supervision based on the target labels encourages the model to learn to close as many as possible gates for background noises where $y_i=0$, than one without this supervision (SG-VAD+AC+$L_{sg}$).

An additional observation from the ablation study is that all models achieve nearly the same validation accuracy, about 98\%. However, the performance on real test sets varies. This observation shows how the proposed method generalizes to unseen domains.

\begin{table}[h!]
  \small{
    \begin{tabular}{|l|c|c|c|}
    \hline
        Configuration & Train Params  & \multicolumn{2}{c|}{AUC-ROC}  \\
    
    &  & AVA & HAVIC\\
    \hline
    \hline
    SG-VAD-R  & 7.8K &  87.9  &  74.9  \\
    SG-VAD + AC & 80.4K & 62.7 & 58.8 \\
    SG-VAD + AC + $L_{sg}$ & 80.4K & 93.2 & 80.8 \\
    \hline
    Proposed Full & 80.4K & 94.3 & 83.3 \\
    \hline
    \end{tabular}
    }
  \caption{\textit{Ablation study results. During inference, each ablation setup's voice activity detection model has 7.8k learned parameters. The second column indicates the number of training parameters. The proposed model, with all of its components, achieves the highest AUC results.}} 
   \label{tab:ablation}
\end{table}

\section{CONCLUSIONS}
\label{sec:conclusions}
This work proposes a novel SG-VAD model for voice activity detection. Our model comprises two networks, the first acts as a dynamic feature selection model trained to select features that contain the speech signal. The second network is a convolution-based classifier that predicts the speech label for each segment. Both networks are trained simultaneously to minimize the sum of two loss terms. We use the feature selection module as our voice activity detector. Our method achieves state-of-the-art results on the AVA-speech evaluation dataset while reducing the size of the prediction netwrok. We further evaluate the importance of our model's components through a series of ablation studies. 
\balance
\bibliographystyle{IEEEbib}
\bibliography{refs}

\begin{thebibliography}{10}

\bibitem{kopuklu2022resectnet}
Okan K{\"o}p{\"u}kl{\"u} and Maja Taseska,
\newblock ``Resectnet: An efficient architecture for voice activity detection
  on mobile devices,''
\newblock {\em Proc. Interspeech 2022}, pp. 5363--5367, 2022.

\bibitem{jia2021marblenet}
Fei Jia, Somshubra Majumdar, and Boris Ginsburg,
\newblock ``Marblenet: Deep 1d time-channel separable convolutional neural
  network for voice activity detection,''
\newblock in {\em ICASSP 2021-2021 IEEE International Conference on Acoustics,
  Speech and Signal Processing (ICASSP)}. IEEE, 2021, pp. 6818--6822.

\bibitem{rho2022vad}
Daniel Rho, Jinhyeok Park, and Jong~Hwan Ko,
\newblock ``Nas-vad: Neural architecture search for voice activity detection,''
\newblock {\em arXiv preprint arXiv:2201.09032}, 2022.

\bibitem{braun2021training}
Sebastian Braun and Ivan Tashev,
\newblock ``On training targets for noise-robust voice activity detection,''
\newblock in {\em 2021 29th European Signal Processing Conference (EUSIPCO)}.
  IEEE, 2021, pp. 421--425.

\bibitem{xu2021lightweight}
Xuenan Xu, Heinrich Dinkel, Mengyue Wu, and Kai Yu,
\newblock ``A lightweight framework for online voice activity detection in the
  wild.,''
\newblock in {\em Interspeech}, 2021, pp. 371--375.

\bibitem{xiong2021computationally}
Yan Xiong, Visar Berisha, and Chaitali Chakrabarti,
\newblock ``Computationally-efficient voice activity detection based on deep
  neural networks,''
\newblock in {\em 2021 IEEE Workshop on Signal Processing Systems (SiPS)}.
  IEEE, 2021, pp. 64--69.

\bibitem{kim2022ada}
Taesoo Kim, Jiho Chang, and Jong~Hwan Ko,
\newblock ``Ada-vad: Unpaired adversarial domain adaptation for noise-robust
  voice activity detection,''
\newblock in {\em ICASSP 2022-2022 IEEE International Conference on Acoustics,
  Speech and Signal Processing (ICASSP)}. IEEE, 2022, pp. 7327--7331.

\bibitem{xu2020polishing}
Tianjiao Xu, Hui Zhang, and Xueliang Zhang,
\newblock ``Polishing the classical likelihood ratio test by supervised
  learning for voice activity detection.,''
\newblock in {\em INTERSPEECH}, 2020, pp. 3675--3679.

\bibitem{makishima2021enrollment}
Naoki Makishima, Mana Ihori, Tomohiro Tanaka, Akihiko Takashima, Shota
  Orihashi, and Ryo Masumura,
\newblock ``Enrollment-less training for personalized voice activity
  detection,''
\newblock {\em arXiv preprint arXiv:2106.12132}, 2021.

\bibitem{chen2020voice}
Yefei Chen, Heinrich Dinkel, Mengyue Wu, and Kai Yu,
\newblock ``Voice activity detection in the wild via weakly supervised sound
  event detection.,''
\newblock in {\em INTERSPEECH}, 2020, pp. 3665--3669.

\bibitem{lee2020dual}
Joohyung Lee, Youngmoon Jung, and Hoirin Kim,
\newblock ``Dual attention in time and frequency domain for voice activity
  detection,''
\newblock {\em arXiv preprint arXiv:2003.12266}, 2020.

\bibitem{li2022voice}
Shu Li, Ye~Li, Tao Feng, Jinze Shi, and Peng Zhang,
\newblock ``Voice activity detection using a local-global attention model,''
\newblock {\em Applied Acoustics}, vol. 195, pp. 108802, 2022.

\bibitem{lindenbaum2021differentiable}
Ofir Lindenbaum, Uri Shaham, Erez Peterfreund, Jonathan Svirsky, Nicolas Casey,
  and Yuval Kluger,
\newblock ``Differentiable unsupervised feature selection based on a gated
  laplacian,''
\newblock {\em Advances in Neural Information Processing Systems}, vol. 34, pp.
  1530--1542, 2021.

\bibitem{yang2022locally}
Junchen Yang, Ofir Lindenbaum, and Yuval Kluger,
\newblock ``Locally sparse neural networks for tabular biomedical data,''
\newblock in {\em International Conference on Machine Learning}. PMLR, 2022,
  pp. 25123--25153.

\bibitem{yamada2020feature}
Yutaro Yamada, Ofir Lindenbaum, Sahand Negahban, and Yuval Kluger,
\newblock ``Feature selection using stochastic gates,''
\newblock in {\em International Conference on Machine Learning}. PMLR, 2020,
  pp. 10648--10659.

\bibitem{miller2017reducing}
Andrew Miller, Nick Foti, Alexander D'Amour, and Ryan~P Adams,
\newblock ``Reducing reparameterization gradient variance,''
\newblock {\em Advances in Neural Information Processing Systems}, vol. 30,
  2017.

\bibitem{figurnov2018implicit}
Mikhail Figurnov, Shakir Mohamed, and Andriy Mnih,
\newblock ``Implicit reparameterization gradients,''
\newblock {\em Advances in neural information processing systems}, vol. 31,
  2018.

\bibitem{chaudhuri2018ava}
Sourish Chaudhuri, Joseph Roth, Daniel~PW Ellis, Andrew Gallagher, Liat Kaver,
  Radhika Marvin, Caroline Pantofaru, Nathan Reale, Loretta~Guarino Reid, Kevin
  Wilson, et~al.,
\newblock ``Ava-speech: A densely labeled dataset of speech activity in
  movies,''
\newblock {\em arXiv preprint arXiv:1808.00606}, 2018.

\bibitem{strassel2012creating}
Stephanie Strassel, Amanda Morris, Jonathan~G Fiscus, Christopher Caruso,
  Haejoong Lee, Paul Over, James Fiumara, Barbara Shaw, Brian Antonishek, and
  Martial Michel,
\newblock ``Creating havic: Heterogeneous audio visual internet collection,''
\newblock in {\em Proceedings of the Eighth International Conference on
  Language Resources and Evaluation (LREC'12)}, 2012, pp. 2573--2577.

\bibitem{warden2018speech}
Pete Warden,
\newblock ``Speech commands: A dataset for limited-vocabulary speech
  recognition,''
\newblock {\em arXiv preprint arXiv:1804.03209}, 2018.

\bibitem{font2013freesound}
Frederic Font, Gerard Roma, and Xavier Serra,
\newblock ``Freesound technical demo,''
\newblock in {\em Proceedings of the 21st ACM international conference on
  Multimedia}, 2013, pp. 411--412.

\bibitem{kuchaiev2019nemo}
Oleksii Kuchaiev, Jason Li, Huyen Nguyen, Oleksii Hrinchuk, Ryan Leary, Boris
  Ginsburg, Samuel Kriman, Stanislav Beliaev, Vitaly Lavrukhin, Jack Cook,
  et~al.,
\newblock ``Nemo: a toolkit for building ai applications using neural
  modules,''
\newblock {\em arXiv preprint arXiv:1909.09577}, 2019.

\bibitem{reddy2020interspeech}
Chandan~KA Reddy, Vishak Gopal, Ross Cutler, Ebrahim Beyrami, Roger Cheng,
  Harishchandra Dubey, Sergiy Matusevych, Robert Aichner, Ashkan Aazami,
  Sebastian Braun, et~al.,
\newblock ``The interspeech 2020 deep noise suppression challenge: Datasets,
  subjective testing framework, and challenge results,''
\newblock {\em arXiv preprint arXiv:2005.13981}, 2020.

\bibitem{park2019specaugment}
Daniel~S Park, William Chan, Yu~Zhang, Chung-Cheng Chiu, Barret Zoph, Ekin~D
  Cubuk, and Quoc~V Le,
\newblock ``Specaugment: A simple data augmentation method for automatic speech
  recognition,''
\newblock {\em arXiv preprint arXiv:1904.08779}, 2019.

\bibitem{devries2017improved}
Terrance DeVries and Graham~W Taylor,
\newblock ``Improved regularization of convolutional neural networks with
  cutout,''
\newblock {\em arXiv preprint arXiv:1708.04552}, 2017.

\bibitem{he2019bag}
Tong He, Zhi Zhang, Hang Zhang, Zhongyue Zhang, Junyuan Xie, and Mu~Li,
\newblock ``Bag of tricks for image classification with convolutional neural
  networks,''
\newblock in {\em Proceedings of the IEEE/CVF Conference on Computer Vision and
  Pattern Recognition}, 2019, pp. 558--567.

\bibitem{ardila2019common}
Rosana Ardila, Megan Branson, Kelly Davis, Michael Henretty, Michael Kohler,
  Josh Meyer, Reuben Morais, Lindsay Saunders, Francis~M Tyers, and Gregor
  Weber,
\newblock ``Common voice: A massively-multilingual speech corpus,''
\newblock {\em arXiv preprint arXiv:1912.06670}, 2019.

\bibitem{reddy2021interspeech}
Chandan~KA Reddy, Harishchandra Dubey, Kazuhito Koishida, Arun Nair, Vishak
  Gopal, Ross Cutler, Sebastian Braun, Hannes Gamper, Robert Aichner, and
  Sriram Srinivasan,
\newblock ``Interspeech 2021 deep noise suppression challenge,''
\newblock {\em arXiv preprint arXiv:2101.01902}, 2021.

\bibitem{gemmeke2017audio}
Jort~F Gemmeke, Daniel~PW Ellis, Dylan Freedman, Aren Jansen, Wade Lawrence,
  R~Channing Moore, Manoj Plakal, and Marvin Ritter,
\newblock ``Audio set: An ontology and human-labeled dataset for audio
  events,''
\newblock in {\em 2017 IEEE international conference on acoustics, speech and
  signal processing (ICASSP)}. IEEE, 2017, pp. 776--780.

\bibitem{zue1990speech}
Victor Zue, Stephanie Seneff, and James Glass,
\newblock ``Speech database development at mit: Timit and beyond,''
\newblock {\em Speech communication}, vol. 9, no. 4, pp. 351--356, 1990.

\end{thebibliography}
\end{document}